\documentclass[aps,twocolumn,groupedaddress]{revtex4-1}
\usepackage{graphicx}
\usepackage{amssymb}
\usepackage{multirow}
\usepackage{subfigure}
\usepackage{tabularx}

\begin{document}

\title{Self-assembly of Chiral Tubules}

\author{Shengfeng Cheng}
\altaffiliation{Present address: Department of Physics, Virginia Polytechnic Institute and State University, Blacksburg, Virginia 24061, USA}
\email{chengsf@vt.edu}
\affiliation{Sandia National Laboratories, Albuquerque, NM 87185, USA}
\author{Mark J. Stevens}
\email{msteve@sandia.gov}
\affiliation{Sandia National Laboratories, Albuquerque, NM 87185, USA}

\date{\today}

\begin{abstract}
The efficient and controlled assembly of complex structures from
macromolecular building blocks is a critical open question in both biological systems
and nanoscience.
Using molecular dynamics simulations 
we study the self-assembly of tubular structures from 
model macromolecular monomers with multiple binding sites on their surfaces
[Cheng et al., Soft Matter, 2012, {\bf 8}, 5666-5678].
In this work we add chirality to the model monomer and a lock-and-key interaction.
The self-assembly of free monomers into tubules yields a pitch value that often
does not match the chirality of the monomer (including achiral monomers).
We show that this mismatch occurs because of a twist deformation that
brings the lateral interaction sites into alignment when the tubule pitch differs from the monomer chirality.
The energy cost for this deformation is small as the energy distributions
substantially overlap for small differences in the pitch and chirality.
In order to control the tubule pitch by preventing the twist deformation, the
interaction between the vertical surfaces must be increased without resulting in
kinetically trapped structures.
For this purpose, we employ the lock-and-key interactions and obtain good
control of the self-assembled tubule pitch.
These results explain some fundamental features of microtubules.
The vertical interaction strength is larger than the lateral in microtubules 
because this yields a controlled assembly of tubules with the proper pitch.
We also generally find that the control of the assembly into tubules is difficult,
which explains the wide range of pitch and protofilament number observed in microtubule assembly.
\end{abstract}

\maketitle

\section{Introduction}

The self-assembly of macromolecular building blocks into structures of 
well-defined shapes and sizes is a fundamental challenge in nanoscience.\cite{MannNM09}
Part of the promise of nanoscience is the development of sophisticated
supramolecular assemblies that possess multifunctional properties and behavior beyond
the capabilities of simpler molecules.\cite{KotovSci10}
Biology is full of examples of such super-structures.
In particular, microtubules are biopolymers that possess features distinct from standard
synthetic polymers because the monomeric building block is a complex protein called tubulin.\cite{Alberts02}
A major feature of microtubules not available in synthetic systems is that they are the track upon which kinesin/dynein motor proteins walk.
Fast depolymerization of microtubules is an essential part of their biological function and quite distinct from 
the typical behavior of synthetic polymers.
These special functions and properties originate from the structural arrangement of tubulin in microtubules, which produces very stiff, chiral tubules.
Understanding how the interactions between tubulin monomers yield the self-assembly
of microtubules and determine their properties is a fundamental open issue.
This issue is pertinent to both biology and materials science, as the special properties of microtubules have led them to be used in synthetic systems to create new materials.\cite{BachandNL04,DootSM07,Hess11,Tamura11}
These studies have stimulated interests in developing supramolecular systems that act as ``artificial microtubules'', i.e., possess features similar or analogous to microtubules.
A far-reaching issue inherent to nanoscience 
is the determination of the essential features of a monomer 
that self-assembles into a tubular structure mimicking microtubules
as well as possessing other features of microtubules. 

Microtubules are both a motivation and an inspiration for our work. 
That is, we want to understand both the assembly of tubulin into microtubules
and the more general principles of the assembly
of artificial tubular structures.
The large amount of data on microtubules helps
define our models and provides something to compare to our simulation results.
The monomer of microtubules is a dimer of two proteins, $\alpha$ and $\beta$-tubulin.
The dimers form protofilaments via noncovalent longitudinal bonding,
and typically 13 protofilaments bind laterally and form tubules 
with an outer diameter $\sim 25$ nm.\cite{Desai97}
Variations in the number of protofilaments and the helical pitch of microtubules assembled from free tubulin {\it in vitro} are significant.\cite{Chretien91, Chretien00}
The structural variations have also been observed to occur to a smaller extent for {\em in vivo} microtubules, where other molecules help control assembly.
These results suggest that the controlled formation of tubules is not simple and understanding the limitations is important for designing synthetic mimics of microtubules.

Tubular structures have been observed in synthetic systems.
Recently, a few cases of supramolecular systems have been developed 
that self-assemble into tubules.\cite{Tarabout11,WangJACS11,WangJACS13} 
Tarabout {\em et al.} constructed a wedge-shaped nanoparticle 
from beta-sheet-forming polypeptides including an artificial peptide.\cite{Tarabout11}
The nanoparticle consists of two layers composed of beta sheets,
which form a bilayer structure with the hydrophilic groups on 
its top and bottom that sandwich the hydrophobic groups.
The top layer is wider than the bottom layer due to larger aromatic residues 
(including artificial peptides) at the edge of the top layer.
They showed that the tubule diameter can be controlled by chemical modifications of an aromatic residue
involved in the contact of nanoparticles.
Wang {\em et al.} have found conditions where polypeptide-grafted comblike polymers form tubular structures.\cite{WangJACS11}
Moreover, they found that gold nanoparticles with grafted poly-(L-glutamic acid) 
can form tubules under certain circumstances.\cite{WangJACS11,WangJACS13}
In these systems, the monomer is effectively a single nanoparticle and 
the nanoparticle-nanoparticle interactions dictate the tubule formation.
It is this class of synthetic tubular systems that we are interested in and have developed some new understanding of 
the role of the interactions between nanoparticles and the control of the self-assembled tubular structures.

There is also a class of small amphiphilic molecules that form tubules typically
by initially forming sheets that subsequently roll into tubules.\cite{SchnurSci93,VautheyPNAS02,Shimizu05,Shimizu08,ForterreSci11,GuoACSNano12}
This class of tubular structures is distinct from microtubules (e.g., not as stiff) and 
is not the focus of this work, although there is some overlap in the underlying assembly phenomena.
For example, Shimizu's group developed wedge-shaped amphiphiles 
by covalently linking hydrophilic groups of different size to the two ends of a hydrophobic spacer.\cite{Masuda01} 
They experimentally showed that they can control the inner diameter of multilayered nanotubes by varying the length of the hydrophobic spacer.\cite{Masuda04}
The wedge geometry does promote tubular structures, but developing these systems into artificial microtubules appears unlikely.

We have previously developed a coarse-grained model monomer with a wedge shape that can self-assemble into
tubular structures with the appropriate binding interactions between monomers.\cite{Cheng12} 
The monomer has lateral interactions that promote ring formation and
vertical interactions that promote filament formation.
We used molecular dynamics (MD) simulations to study the self-assembly process and obtained 
a diagram of the self-assembled structures as a function of lateral and vertical interaction strengths.
Since our model monomers can self-assemble into tubular structures 
and the resulting tubules exhibit similar structural polymorphism as microtubules, 
we have a model that can be used to systematically examine 
the physical origin of structural variations and
explore ways to control the structure.
Furthermore, the strength of lateral and vertical bonds between tubulin monomers
clearly plays important roles in determining the structure of microtubules,
but their effects are hard to probe experimentally since
natural evolution only leads to one particular set of interaction strengths 
between tubulin.\cite{VanBurenPNAS02}
Directly measuring the interactions between monomers is typically not possible
because experiments usually only determine the net free energy difference that is the sum of many interactions.
Our model system offers an opportunity to directly calculate the energetics and the structural dynamics of tubules in more detail than previously attempted.

The focus of this work is the physical origin of the factors controlling tubule assembly.
To this end, we have added new features to our model monomer to improve 
the structural control of the self-assembled tubules 
based on new understanding of the physical nature of the assembly process.
In particular, we focus on controlling the helicity of self-assembled tubules and have
added an explicit chirality to the monomer so that the desired tubule helicity can be input from the monomer chirality.
Unexpectedly, our earlier simulations produced helical tubules but with achiral monomers.\cite{Cheng12}
We show here that this tubule helicity is a consequence of an underlying rotational symmetry that can occur through a twist deformation of protofilaments.
Twist of tubules also occurs with chiral monomers and 
enables a range of tubule helicity about the desired value to occur, i.e., the monomer chirality.
In other words, mismatch occurs between the
chirality of building blocks and the pitch of assembled tubules.
We incorporate a lock-and-key binding mechanism for the vertical interactions between monomers
to more precisely constrain the tubule helicity. 
Our studies show that the lock-and-key binding combined with monomers
with built-in chirality can be used
to achieve structural control in the self-assembly of tubules,
and therefore provide a guidance on the design of building blocks
that will efficiently self-assemble into controlled tubular structures.

The remainder of this paper is organized as follows.
The simulation methods are briefly described in the next section.
In the Results and Discussion section, we first examine the energetics of the achiral system and
explain the physical origin of the helical tubules that assemble from the achiral wedges.
We then include explicit chirality in the monomer design and present the assembly and 
energetics for the chiral systems.
Finally, we discuss the results of the systems that incorporate both chirality and
a lock-and-key interaction for the vertical binding. These systems provide the best control of the assembly of tubular structures.
Conclusions are included in the last section.

\section{Simulation Methods}

\begin{figure}[tb]
\centering
\includegraphics[width=3in]{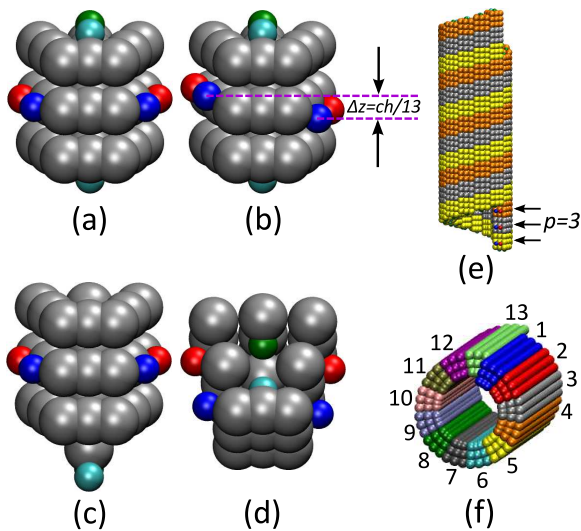}
\caption{(a) An achiral wedge monomer $M_0$ designed for nonhelical tubules 
(i.e., pitch $p=0$). 
(b) A chiral wedge monomer $M_2$ with chirality $c=2$ (i.e., designed for
tubules with $p=2$).
The chirality $c$ determines via $\Delta z=ch/13$
the displacement $\Delta z$ between the binding sites on the left and right sides of the wedge,
where $h=3\sigma$ is the wedge height.
(c) An achiral wedge monomer with a
lock-and-key configuration for the vertical binding, $M_0^{LK}$;
the vertical binding sites stick out at the bottom surface
and are indented at the top surface. 
(d) A chiral wedge monomer with $c=2$ and with a
lock-and-key configuration for the vertical binding, $M_2^{LK}$;
the top view shows that the vertical binding sites are buried below the top surface.
(e) A 13\_3 tubule with $N=13$ protofilaments and pitch $p=3$ (each helix is colored differently).
(f) The same tubule as in (e) but with a top view showing 13 protofilaments 
(each protofilament is colored differently);
for clarity the binding sites are not shown in (f).
}
\label{Fig_buildingblock}
\end{figure}

The wedge-shaped building blocks are shown in Fig.~\ref{Fig_buildingblock}. Each monomer
is treated as a rigid body in simulations reported here.
The core of the monomer is a wedge
deformed from a cube composed of 27 particles (gray spheres in Fig.~\ref{Fig_buildingblock}) 
with an undeformed lattice constant $1\sigma$, where $\sigma$ is the particle diameter. 
The deformation is imposed in such a way that 13 wedges will fit to form a closed ring.
The interaction between these core sites on two monomers is
modeled as a short-range repulsion using the Lennard-Jones 12-6 potential 
$U_{\rm LJ}(r)=4\epsilon \left[ (\sigma/r)^{12}-(\sigma/r)^6\right]$,
where $\epsilon$ is the energy scale.
This potential is cut off and shifted to 0 at $r_c = 1.0\sigma$.
The short-range repulsion models the excluded volume interaction between monomers.
The temperature $T$ is set as $1.0\epsilon/k_{\rm B}$ 
($k_{\rm B}$ is the Boltzmann constant) in our simulations.
We will use $k_{\rm B}T$ as the energy unit.
Attractions between monomers occur through
8 binding sites (colored spheres in Fig.~\ref{Fig_buildingblock})
placed on the lateral and vertical faces of the wedge. 
A binding site on the left (top) face of a wedge only interacts with the binding site in the same color 
and on the right (bottom) face 
of another wedge and vice versa.
The bonding interaction is modeled as 
a soft potential $U_{\rm B}(r)=-A\left[ 1+{\rm cos}(\pi r / r_a)\right]$,
where $A$ is the binding strength and $r_a$ the interaction range. 
The well depth of this potential is $2A$ at $r=0$.
We use $A_L$ and $A_V$ (in the unit of $k_{\rm B}T$) to designate the strength of the
lateral and vertical binding interactions, respectively.
We use this potential form because it smoothly goes to zero at $r=r_a$.
In our previous work and this paper, we fix $r_a=1.0\sigma$.
The potential form allows for variation of $r_a$, but we leave that for future work.
Since $U_{\rm B}(r)$ is isotropic, a minimum of 2 binding sites on each face
is needed to break the rotational symmetry when two monomers bind,
which is crucial to ensure that a pair of bonded monomers
have the proper orientation.

In the original design, the lateral binding sites are placed at $0.5\sigma$ outside the lateral face
and in the middle plane of the core particles, as shown in Fig.~\ref{Fig_buildingblock}(a). 
The vertical binding sites are similarly placed with respect to the top or bottom face.
Such a building block has a mirror symmetry and 
its achiral geometry is designed for nonhelical tubules.
This monomer is designated as $M_0$.
In this work, we extend the wedge model to treat chiral monomers.
Chirality is introduced by shifting the lateral binding sites
oppositely on the left and right side of the wedge, respectively.
In general, a chiral monomer will de designated as $M_c$, where
$c$ designates the chirality of the monomer. 
For a $M_c$ monomer, the displacement between the lateral binding sites is $\Delta z \equiv ch/13$,
where $h=3\sigma$ is the wedge height.
The $M_2$ monomer is shown in Fig.~\ref{Fig_buildingblock}(b).
We treat cases where the $M_c$ monomers are designed to assemble into tubules with pitch $p=c$.
To label tubular structures, we follow the literature and use $N\_p$ to denote tubules with $N$ 
protofilaments and pitch $p$ (counted in the unit of building blocks and also called 
{\it helix start number} in the literature on microtubules).\cite{Chretien00}
An illustration of $p$ and $N$ is given in Figs.~\ref{Fig_buildingblock}(e) and (f).

One surprise found in our previous simulations using the achiral $M_0$
is the preferential assembly of helical tubules with pitch 1 or
even 2.\cite{Cheng12}
New results here show that tubules with pitch mismatching the monomer chirality
still assemble even for chiral monomers.
We found that one factor that can help suppress the mismatch
and better control the tubule pitch
is to make vertical binding stronger.
Since simply increasing $A_V$ leads to kinetically trapped clusters of wedges, we instead explore
a lock-and-key mechanism for the vertical binding interactions, which is introduced by modifying
the location of the vertical binding sites.
These sites together with the central line of core particles
are displaced vertically by $0.75\sigma$, so that
the top pair of binding sites are
buried below the top surface of the wedge by $0.25\sigma$,
while the bottom pair stick out
of the bottom surface by $1.25\sigma$.
The lock-and-key modification and chirality are combined
to make building blocks labeled $M_c^{LK}$, 
where the superscript $LK$ stands for lock-and-key. 
$M_0^{LK}$ and $M_2^{LK}$ are shown in Figs.~\ref{Fig_buildingblock}(c)
and \ref{Fig_buildingblock}(d), respectively.

All simulations were performed with the LAMMPS simulation package.
The equations of motion were integrated using a velocity-Verlet algorithm
with a time step $\delta t = 0.005\tau$,
where $\tau \equiv \sigma (m/\epsilon)^{1/2}$ is 
the unit of time and $m$ the mass of one site.
The simulations studying the self-assembly of various
monomers involved 1000 wedges and were run for $4\times 10^6\tau$
to $8\times 10^6\tau$.
The initial state was a low-density gas of monomers uniformly distributed
in the simulation box.
The temperature of the systems was kept at $1.0\epsilon/k_{\rm B}T$
with a Langevin thermostat of damping rate $1.0\tau^{-1}$.

\section{Results and Discussion}

\subsection{Helicity of Tubules Assembled from Achiral Wedges}

\begin{figure}[tb]
\centering
\includegraphics[width=2.75in]{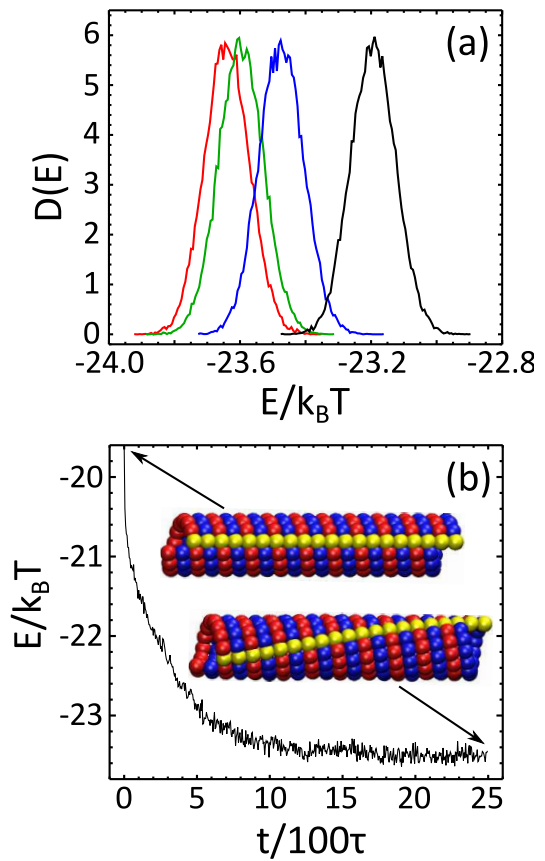}
\caption{(a) The probability density of energy distribution per monomer, $D(E)$,  for 
equilibrated $13\_p$ tubules with $p=0$ (red), 1 (green), 2 (blue), and 3 (black),
from left to right.
(b) A 13\_2 tubule starting with straight protofilaments evolves into
a lower-energy state with twisted protofilaments.
In the images each wedge making up the tubule is shown as a single sphere. 
The helices formed via the lateral binding are shown in red and blue, respectively, and
one protofilament is colored yellow to emphasize the twist transformation.
Both (a) and (b) are for $A_L=4.2$ and $A_V=2.6$.
}
\label{Fig_twist}
\end{figure}

The first important issue to be addressed 
is why achiral monomers form helical tubules.
To this end, we built $13\_p$ tubules from $M_0$ monomers with $p$ ranging from 0 to 3.
The tubules start in a state with straight protofilaments. 
The energy distribution densities per monomer, $D(E)$, 
calculated for each $p$ from MD simulations
after the tubules reach equilibrium are shown in Fig.~\ref{Fig_twist}(a).
As expected, the 13\_0 tubule has the lowest mean energy.
However, the energy distribution of the 13\_1 tubule has a large overlap with that of the 13\_0 tubule.
Even the energy distribution of the 13\_2 tubule overlaps with that of the 13\_0 tubule.
Being so close energetically explains
the frequent formation of 13\_1 and 13\_2 tubules
by the achiral $M_0$ monomers in the assembly simulations starting with free monomers
(i.e., the initial condition is a ``gas'' phase of wedge monomers).
The overlap becomes negligible as $p$ increases to 3, which explains the lack of 13\_3 tubules in the self-assembly products.
While the energy difference per monomer at the peaks of $D(E)$ for 13\_0 and 13\_3 is only about 0.4 $k_{\rm B} T$, the difference between the initial tubules formed in the self-assembly simulations is much larger since many monomers are involved in the nucleation and the total energy difference depends on the initial tubule size.
A single turn of 13 monomers puts the total energy difference above 5 $k_{\rm B} T$. 
In some cases ($A_V>A_L$) multiple partial turns form before the tubule state occurs, and the energy difference will be consequently much larger than 5 $k_{\rm B} T$. 

\begin{figure*}[bt]
\centering
\includegraphics[width=6in]{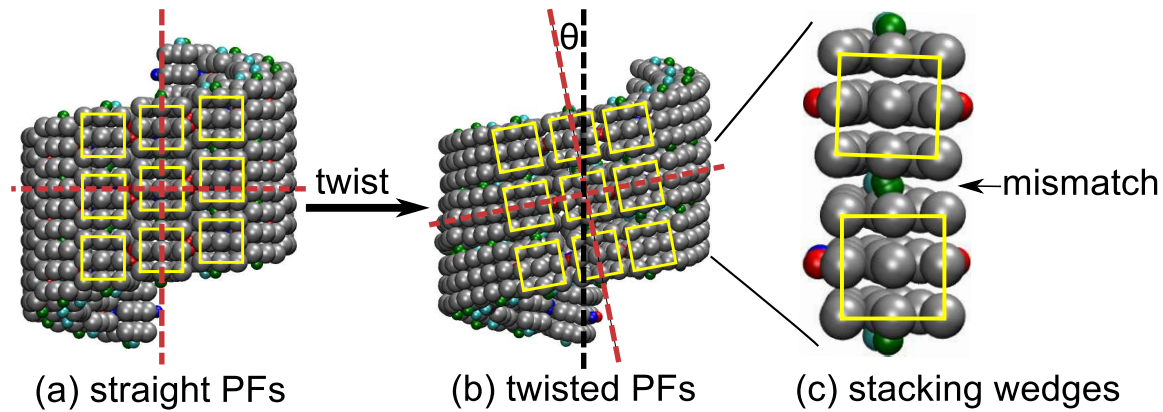}
\caption{A 13\_2 tubule starting with straight protofilaments (a) undergoes
a skew deformation to transform into a tubule with
twisted protofilaments (b) at $A_L=4.2$ and $A_V=2.6$,
which results in a better packing of wedges.
For clarity only a short part of the whole tubule is shown.
Twist angle $\theta$ is the angle between
protofilaments and the central axis of the tubule.
(c) The stacking of two consecutive wedges in a twisted protofilament
involves an offset between the binding sites at vertical interfaces
due to the rotation about the central axis.
}
\label{Fig_angle_illustration}
\end{figure*}

A structural transformation must occur for the $p=1$ or 2 tubules to have energies
that overlap with $p=0$ 
because helical tubules (i.e., $p>0$) with straight protofilaments have the neighboring lateral attractive sites displaced resulting in higher energies.
To probe the role of twist deformation, 
we examined the energy and geometry for the 13\_2 tubule.
Starting from straight protofilaments as shown in the top image of Fig.~\ref{Fig_twist}(b),
this tubule quickly transforms into 
a steady state with twisted protofilaments as shown in the bottom image of Fig.~\ref{Fig_twist}(b).
During the transformation, the energy per monomer
decreases with time by about 4 $k_{\rm B}T$ as shown in Fig.~\ref{Fig_twist}(b).
The energy decreases as protofilaments get twisted because
the twist deformation brings the lateral interaction sites into alignment.
The protofilaments in the nonhelical 13\_0 tubule do remain straight as expected.

Figure \ref{Fig_angle_illustration} illustrates the twist transformation at the monomer level
with the full geometry and local rotation of wedges visualized, 
showing how twisted protofilaments can help lower the energy of a helical tubule made out of $M_0$ monomers. 
As expected for straight protofilaments in a 13\_2 tubule (Fig.~\ref{Fig_angle_illustration}(a) dotted red line), the lateral binding is clearly not maximized because of the displacement between the lateral attractive sites.
As shown by the dotted red lines in Fig.~\ref{Fig_angle_illustration}(b), in the twist transformation the rotation of monomers 
aligns the lateral sites bringing their binding strengths close to their maxima.
Consequently, the total energy of a helical tubule is reduced, which stabilizes the tubule.
Thus the energy of tubules with $p\ne c$ can be close to the preferred $p=c$ case because
the twist deformation of the protofilaments substantially alters the energy
by making the lateral alignment closer to the ideal configuration.

A more detailed examination of the packing of $M_0$ wedges in helical tubules reveals the limit of twist deformation.
In a twisted protofilament, the wedges must rotate about their vertical axis 
so that its inner surface always points to the interior of the tubule, which
introduces an offset between the vertical binding sites
of two consecutive stacking wedges, as shown in Fig.~\ref{Fig_angle_illustration}(c).
For $M_0$ monomers, this offset is very small in a $13\_1$ tubule, which
makes it energetically close to the 13\_0 tubule.
However, as the pitch of a helical tubule gets larger,
the offset increases as a result of the increasing amount of twist.
The offset is clearly visible for a 13\_2 tubule as in Fig.~\ref{Fig_angle_illustration}(c)
and even more dramatic for a 13\_3 tubule (see Fig.~S1 in the Supplement).
For this reason, the energy distributions for the 13\_2 and 13\_3 tubules shift to higher 
and higher values, and the 13\_3 distribution is well separated from the 13\_0 one. 
Only so much twist can occur without resulting in an expensive mismatch between the vertical binding pairs.

The result of multiple pitch values occurring for tubules in thermal equilibrium reveals a limit of treating a microtubule
as an elastic tubule within the framework of continuum elasticity,
where the twist of the tubule always costs energy and induces a restoring force.\cite{Hunyadi05,PampaloniPNAS06,Hunyadi07,SeptPRL10} 
In our simulations, the continuum theory breaks down because of the discrete nature of the building blocks.
The monomers are not deformed in the twist deformation, and thus there is no internal elastic cost.
All the interactions are between the surfaces, and the lateral and vertical sites play distinct roles with
the energetic cost occurring primarily at the vertical contacts.
While the lowest energy state is the same in either treatment (i.e., the tubule with pitch that matches the monomer chirality has the lowest energy), the simulations with discrete building blocks
reveal multiple states in equilibrium and the role of the surface interactions between the monomers.

\begin{figure}[tb]
\centering
\includegraphics[width=2.75in]{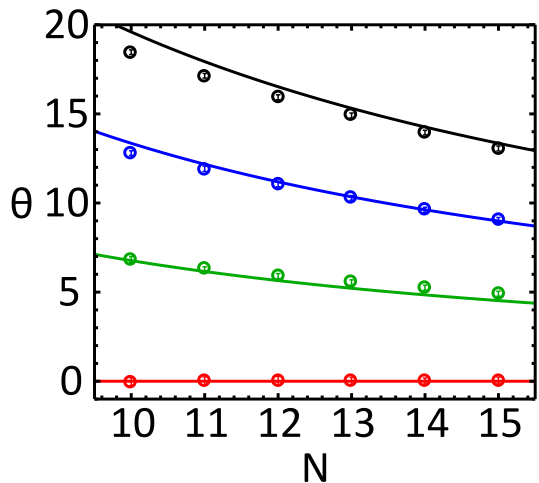}
\caption{Twist angle $\theta$ of tubules
formed by $M_0$ monomers and 
with various number ($N$) of protofilaments
and pitch values: $p=0$ (red), $p=1$ (green), 
$p=2$ (blue), and $p=3$ (black)
at $A_L=3.0$ and $A_V=4.2$.
Symbols are simulation results 
with error bars comparable or smaller than the symbol size.
Lines represent the corresponding 
predictions of Eq.~(\ref{Eq_LAM}) with $h=3\sigma$ and $w=2.53\sigma$.
}
\label{Fig_angle_S0}
\end{figure}

The twist deformation of protofilaments can be quantified by a twist angle $\theta$, which
is the angle between the protofilaments and the central axis of the tubule (Fig.~\ref{Fig_angle_illustration}(b)).
For a $N\_p$ tubule, $\theta$ depends
on both $N$, $p$, and $c$, and
their relation can be easily derived 
through geometrical considerations
based on the packing of anisotropic objects
on the curved surface of a tubule,
called the lattice accommodation model in the literature.\cite{Chretien00}
In general, for a $N\_p$ tubule built from $M_c$ monomers,
the twist angle $\theta$ is given by
\begin{equation}
{\rm tan}\theta = \frac{h}{w} \left( \frac{p}{N} -\frac{c}{N_0} \right),
\label{Eq_LAM}
\end{equation}
where $h$ and $w$ are the height and width of the wedge,
respectively, and $N_0=13$ is the designed number
of protofilaments in an ideal tubule. 
Values of $\theta$ for tubules assembled from $M_0$
are shown in Fig.~\ref{Fig_angle_S0}.
In this case, Eq.~(\ref{Eq_LAM}) indicates that $\theta$ is always 0 for tubules with $p=0$, 
no matter how many protofilaments the tubule contains.
The results for $N\_0$ tubules in Fig.~\ref{Fig_angle_S0} is consistent with this prediction.
For $N\_p$ tubules with $p\ne 0$, $\theta$ depends on $N$ via Eq.~(\ref{Eq_LAM}),
as confirmed by the corresponding simulation results in Fig.~\ref{Fig_angle_S0}.
We have also studied the effects on $\theta$ of both $A_L$ and $A_V$.
As expected, $\theta$ is generally insensitive to either $A_L$ or $A_V$ because
$\theta$ is mainly determined by the geometric features
(i.e., width, height, and chirality) of the building blocks as expressed in Eq.~(\ref{Eq_LAM}).
However, one thing to notice is that results in Fig.~\ref{Fig_angle_S0}
are obtained with tubules starting with predetermined pitch
and protofilaments that are appropriately twisted according to Eq.~(\ref{Eq_LAM}).
In this case the pitch does not change during the simulation
and the tubule only fluctuates around the steady configuration, 
which is close to the starting one but generally only metastable.
The situation becomes much more complicated
if we use tubules starting with straight protofilaments,
where the relative strength of $A_V$ and $A_L$ plays an important role
in affecting the stability of the tubules.
More details are given in the Supplement (see Figs.~S2-S4).

\begin{figure}[!Htb]
\centering
\includegraphics[width=2.75in]{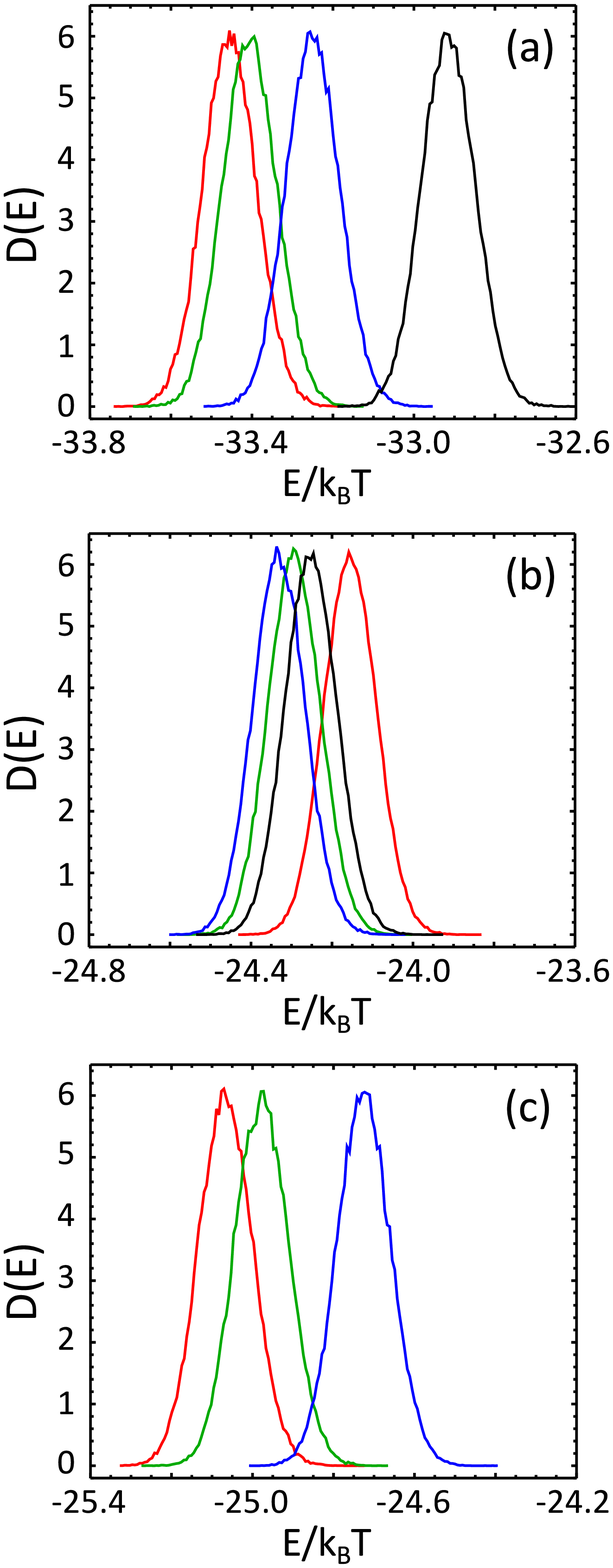}
\caption{The probability density of energy distribution per monomer, $D(E)$, for
various $13\_p$ tubules:  $p=0$ (red), 1 (green), 2 (blue), and 3 (black).
(a) $M_0$ at $A_L=3.0$ and $A_V=6.3$;
(b) $M_2$ at $A_L=3.0$ and $A_V=3.9$;
and (c) $M_0^{LK}$ at $A_L=3.0$ and $A_V=6.3$, where
the 13\_3 tubule is unstable and its $D(E)$ not calculated.
}
\label{Fig_overlap}
\end{figure}

The above results show that the vertical interaction is important in limiting
the range of tubule helicity since it regulates the twist of protofilaments.
Yet, we have found that increasing $A_V$ to increase the energy cost of twist is insufficient to control the helicity of tubules.
Energy distributions of prebuilt tubules with various pitch values 
only show slightly less overlap even for $A_V \gg A_L$.
For example, the energy distributions for the 13\_0 and 13\_1 tubule shown 
in Fig.~\ref{Fig_overlap}(a) for $A_V=6.3$ look very similar to that in Fig.~\ref{Fig_twist} for $A_V=2.6$.
For the 13\_2 tubule the overlap with the energy distribution of the 13\_0 tubule 
is clearly reduced, but some overlap still exists.
In general, the reduction in the energy overlap at large $A_V$ is not sufficient to change the
range of tubule helicity; 
tubules with $p>0$ will still self-assemble for $M_0$ monomers even for very large $A_V$.
Moreover, self-assembly simulations of free monomers with large $A_V$ actually yield kinetically trapped clusters and other defected structures.\cite{Cheng12} 
Thus, these results imply that to control the tubule helicity 
additional features (e.g., interactions) will have to be added to the monomer.

\begin{figure*}[htb]
\centering
\includegraphics[width=6.5in]{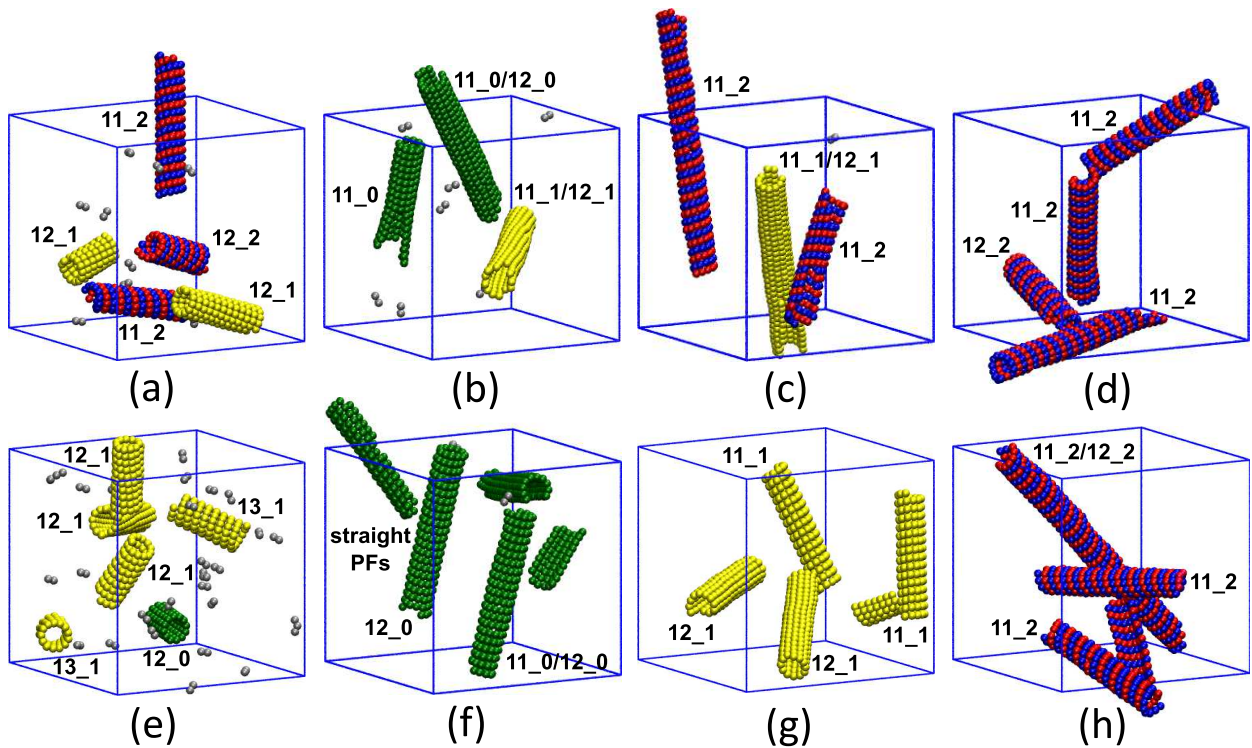}
\caption{The self-assembly of tubules with various monomers:
(a) $M_0$ at $A_L=4.2$ and $A_V=2.6$;
(b) $M_0$ at $A_L=3.0$ and $A_V=3.9$;
(c) $M_1$ at $A_L=3.0$ and $A_V=3.9$;
(d) $M_2$ at $A_L=3.0$ and $A_V=3.9$;
(e) $M_0^{LK}$ at $A_L=4.4$ and $A_V=4.2$;
(f) $M_0^{LK}$ at $A_L=3.0$ and $A_V=6.3$;
(g) $M_1^{LK}$ at $A_L=3.6$ and $A_V=5.4$;
(h) $M_2^{LK}$ at $A_L=3.0$ and $A_V=6.3$.
Each sphere represents a wedge monomer.
The color code is as follows. Structures shown in green are either nonhelical tubules (i.e., $p=0$)
or unclosed sheets with straight protofilaments.
Tubules with $p=1$ are shown in yellow.
Tubules with $p=2$ are shown with helices colored in red and blue, respectively.
Oligomers are shown in silver and free monomers are not included.
}
\label{Fig_tubules}
\end{figure*}

While assembly kinetics is not the focus of this paper, 
it plays an important role in the self-assembly of tubules and a
few important points need to be made.
The self-assembly simulations start with free monomers uniformly distributed 
with a low density in a box.
When $A_L > A_V$, $M_0$ monomers tend to first form rings or helical strands,
which then grow into nonhelical or helical tubules, respectively.\cite{Cheng12}
On the contrary, if $A_V > A_L$, then $M_0$ monomers tend to form
protofilaments first, which then form curved sheets (see examples in Fig.~\ref{Fig_tubules}(f)) that eventually close up into tubules.
This process is similar to the proposed self-assembly pathway of tubulin into microtubules.
\cite{MozziconacciPLoS08}
The final structure of the tubule is strongly affected by the closure event between
the two edges of a curved sheet. 
For $M_0$ monomers, if $A_V$ is only slightly stronger than $A_L$,
then the tubule can still become helical because the sheet is flexible and
thermal fluctuations can introduce an offset between the two edges of the sheet 
when they meet and close up.
If the curved sheets were stiff enough to 
suppress the effects of thermal fluctuations,
then the closure event would be more controlled by the intrinsic chirality of the monomer.
However, for the range of interaction strengths at which wedges do self-assemble into tubules, 
the curved sheets are typically not stiff enough to make thermal fluctuations negligible.

The flexibility of the curved sheets impacts the number of protofilaments of a tubule as well.
Especially at $A_V>A_L$, assembled tubules tend to have protofilament number fewer than the designed value 13. 
Besides the twisting fluctuation in the sheets, 
there are circumferential fluctuations that tend to bring the opposite sides of the sheet into contact and closure when the number of protofilaments is only 11 or 12,
even though the monomer width is chosen to allow
13 wedges to fit from purely geometric considerations. 
After the closure,
it is virtually impossible for other wedges or protofilaments to squeeze into the lattice
to make 13-protofilament tubules. 
In other words, tubules are kinetically trapped in states with protofilaments fewer than designed.
It remains an interesting open question that how 
the design of wedge monomers can be tweaked to address the issue of kinetic trapping
and to promote the formation of tubules containing 13 protofilaments.
One obvious possibility is to squeeze the width of wedges so that
sheets containing fewer than 13 protofilaments are unlikely to close.
This direction will be explored in future work.

The previous discussion on the effects of $A_L$ and $A_V$ on the
tubule twist, tubule energy distributions, and assembly kinetics provides
the basis to understand the self-assembly of systems starting with free monomers,
i.e., the initial state is a gas phase. 
Results on various systems containing 1000 monomers are shown in Fig.~\ref{Fig_tubules}.
Here we just discuss parts (a) and (b), which involve the $M_0$ monomer.
In Fig.~\ref{Fig_tubules}(a) where $A_L=4.2> A_V=2.6$,
all assembled tubules are helical with pitch 1 or 2.
We have performed more than one simulation for these parameters and nonhelical tubules ($p=0$) do form in some cases, 
but this result is indicative of the overall finding that helical tubules are more common 
than nonhelical even though the monomer has no chirality,
as long as $A_L>A_V$.
When the self-assembly is induced at $A_V=3.9>A_L=3.0$ (Fig.~\ref{Fig_tubules}(b)), 
then the pitch is 0 or 1, closer to the monomer chirality.
In both cases $N$ is 11 or 12, which is a result of the closure dynamics discussed above.

\subsection{Self-assembly of Chiral Wedges}

We now discuss results for the self-assembly of chiral monomers.
Chirality is important for better modeling microtubules, which have pitch 3.
Chiral wedges are produced by introducing an offset along the vertical direction
between the two sets of lateral attractive sites on the opposite sides of the wedge.
The amount of offset sets the value of chirality.
As noted earlier, ideally the $M_c$ monomer would yield a tubule with pitch $p=c$.
The $M_2$ monomer is shown in Fig.~\ref{Fig_buildingblock}(b).

We first present the energy distribution of prebuilt tubules with various pitch values. 
An example is shown in Fig.~\ref{Fig_overlap}(b) for the $M_2$ monomer
at $A_L=3.0$ and $A_V=3.9$.
As expected, the $p=2$ tubule now has the lowest mean energy per monomer.
The $p=1$ and 3 tubules have the next two higher average energies per monomer.
Their energy distributions show large overlaps with that of the $p=2$ tubule.
Even the energy distribution of the $p=0$ tubule has some overlap with the rest.
The spread of the four distributions is smaller than for the $M_0$ case in
Fig.~\ref{Fig_overlap}(a).
For the $M_2$ (generally $M_c$) monomer, the protofilaments in tubules with pitch $p\ne 2$ (generally $p\ne c$) are twisted,
similar to those in the helical tubules made of $M_0$ monomers (Fig.~\ref{Fig_twist}(b)).
The twist angle is found to be consistent with the prediction of Eq.~(\ref{Eq_LAM}).
Thus, while including chirality in the monomer shifts the lowest energy state to the tubules with the preferred pitch (as input from the monomer chirality), 
the twist deformation of protofilaments still yields overlapping energy distributions with mismatched pitch values.

Images of the tubules formed during the self-assembly simulations of $M_0$, $M_1$, and $M_2$ at $A_V=3.9>A_L=3.0$ are shown in Figs.~\ref{Fig_tubules}(b)-(d), respectively.
As the value of $c$ shifts so does the pitch $p$ of the assembled tubules.
The results are similar between the $M_0$ and $M_1$ cases.
There are tubules with $p=c$ formed, but there also tubules with other pitch values.
Surprisingly, the self-assembly of $M_2$ monomers shows that
all the assembled tubules have pitch 2, which is the same
as the built-in chirality of the $M_2$ monomers.
Three independent runs with different initial conditions all produce tubules only with $p=2$.
But our expectation is that this is just a consequence of small samples.
For the three cases in Fig.~\ref{Fig_tubules}(b)-(d), when the interactions strengths are switched to stronger lateral binding over vertical ($A_L=4.2>A_V=2.6$), a wide range of pitch values occurs (see Table 1 in the Supplement).
Particularly in the strong $A_L$ limit the tubules with mismatched pitch values seem to dominate,
indicating the effects of strong lateral bonds on the nucleation of seeding structures and the assembly pathway.
More discussion is included at the end of Sec.~\ref{sec_LK}.
Overall, the data indicate that though the pitch of tubules can be varied in a certain range by using chiral monomers, it is difficult to achieve a precise match between the pitch
and chirality. 
The pitch of assembled tubules usually shows a range centered on the chirality of monomers with $|p-c| \le 2$.

\subsection{Self-assembly of Wedges with Lock-and-Key Binding}
\label{sec_LK}

We have found that the twist deformation makes control of the tubule pitch difficult in most cases.
While large $A_V$ would limit or even suppress twist, it generally
leads to kinetically trapped structures rather than well-defined tubules.
An additional feature or modification of the vertical interaction is required
in order to prevent twist.
To this end, we introduce a lock-and-key mechanism 
for the vertical binding. 
The vertical binding sites together with the central line of core sites of the wedge
were shifted along the vertical direction by $0.75\sigma$
such that on the bottom face the attractive sites stick out,
while on the top face the attractive sites become buried 
below the surrounding core sites.
In this way a simple lock-and-key configuration is created.
The vertical lock-and-key (LK) modification on $M_c$ monomers
is labeled $M_c^{LK}$.
Images of the $M_0^{LK}$ and $M_2^{LK}$ monomer are shown in 
Fig.~\ref{Fig_buildingblock}(c) and (d), respectively.

The energy distributions for prebuilt tubules of $M_0^{LK}$ monomers 
are shown in Fig.~\ref{Fig_overlap}(c)
at $A_V=6.3 > A_L=3.0$. These interaction strengths are just strong enough 
to induce self-assembly for this lock-and-key monomer as determined by our assembly simulations starting with free monomers.
In general, the addition of the lock-and-key requires a larger value of $A_V$ for the monomers to bind because extra energy is needed to insert the key and to compensate for the stronger repulsion between the core sites 
(see Fig.~S5 and relevant discussion in the Supplement).
Compared with tubules of monomers without the vertical lock-and-key binding, 
overlaps between the energy distributions for tubules with various pitch values are reduced in the lock-and-key case.
For the $M_0^{LK}$ monomer, while a significant overlap still exists for tubules with $p=1$ and $p=0$, the overlap between $p=2$ and $p=0$ is almost completed gone and tubules with $p=3$ are not even stable anymore.
Thus, tubules with the lowest energy per monomer (e.g., tubules with $p=0$ for $M_0^{LK}$ monomers) will be more favored during the self-assembly of lock-and-key monomers.
The self-assembly of $M_0^{LK}$ monomers with the same interaction strengths as for Fig.~\ref{Fig_overlap}(c) is shown in Fig.~\ref{Fig_tubules}(f).
Again, these simulations start with a gas phase of monomers.
For $M_0^{LK}$, all the assembled tubules are nonhelical with pitch 0, matching the chirality of $M_0^{LK}$.
Some clusters are curved sheets that do not close up into tubules yet because of the exhaustion of free monomers. 
However, in these sheets the protofilaments are straight, and 
if more monomers were supplied to the system,
the tubules would be expected to be nonhelical when they eventually form.

Two examples of the self-assembly of chiral monomers with the lock-and-key geometry 
are shown in Fig.~\ref{Fig_tubules}(g) and (h) for $M_1^{LK}$ and $M_2^{LK}$, respectively.
They show that the lock-and-key mechanism produces a match
between the pitch of tubules and the chirality of monomers. 
This match is better than what we expect from the energy distributions, which imply that mismatched cases should occur. 
We expect that simulations of larger systems that can form many tubules would produce a distribution of pitch values.
Nonetheless, as the comparison between Figs.~\ref{Fig_tubules}(c) and (g) indicates, the addition of the lock-and-key 
mechanism for the vertical binding substantially shifts the assembled structures toward tubules with pitch matching the monomer chirality.
Results of the self-assembly simulations of $M_c^{LK}$ monomers 
at various combinations of $A_L$ and $A_V$ are summarized in Table S2 of the Supplement.
This table shows that there are some ($A_L$, $A_V$) that yield only $p=c$ tubules, but also there are cases with similar ($A_L$, $A_V$) that have $p \ne c$, which is what we expect based on the energy distributions.
Overall, the results are indicative of an improved control of the tubule helicity and demonstrate the basic concept that strong vertical interactions are required in order to limit the amount of protofilament twist and to control the tubule pitch.

The relative importance of vertical interactions can be seen in the comparison between Figs.~\ref{Fig_tubules}(e) and (f) for $A_L > A_V$ and $A_V > A_L$, respectively, for the $M_0^{LK}$ monomer.
For the case with stronger lateral interactions, while the chirality range is narrower and closer to the monomer chirality 
with the lock-and-key binding than without (compare  Figs.~\ref{Fig_tubules}(e) and (a)), 
the most common tubule formed by $M_0^{LK}$ has pitch $p=1$ that still does not match the monomer chirality $c=0$.
Kinetics may play an important role in producing the mismatch. 
For stronger lateral interactions, tubules typically initiate by the formation of rings.
The free energy barrier between forming a nonhelical ring and a helical one is small.
Even if the initial binding yields a nonhelical ring, the single bond between a pair of monomers is weak and fluctuations can break the bond to allow the two ends to slip by each other and then bind together vertically forming a helical ring (see Fig. 9 in Ref. 22).
In this manner, helical tubules tend to form when $A_L > A_V$ even for the $M_0^{LK}$ monomer.
More generally, tubules with $p\ne c$ tend to form more easily for $M_c^{LK}$ monomers at $A_L > A_V$ (see Table S2 in the Supplement).

\section{Conclusions}

We have studied how the interactions and molecular geometry of a simple macromolecular monomer determine the assembly into tubular structures particularly with respect to the chirality of the system.
Our results have important implications in the design of macromolecular building blocks that efficiently self-assemble into tubules.
In order to form tubules with a given pitch value, we introduced
chirality into our model monomers such that in the ideal structure
the tubule pitch would match the monomer chirality.
However, we generally find that the self-assembly from free monomers into tubules with a specified pitch value does not occur.
Instead a range of pitch values emerges.
We show that the twist deformation 
stabilizes tubules with pitch inconsistent with the chirality of monomers
by better aligning neighboring monomers so that they still bind well.
These twisted tubules can have energy distributions that substantially overlap
with that of the untwisted tubule with pitch equal to the monomer chirality.
However, there is a limit to the amount of twist that yields overlapping energy distributions.
Besides rotating monomers about the radial axis and aligning the lateral binding sites, twist also rotates the monomers about their vertical axis, which reduces the alignment of the vertical binding sites.
Thus, to prevent twist a strong vertical interaction is required that makes protofilaments stiffer and strongly raises the cost of vertical misalignment caused by the twist.
A similar trend was found recently in the packing of filament bundles, where the ground state of a bundle of stiff filaments
tend to be untwisted, while filaments with low stiffness form twisted bundles.\cite{BrussPNAS12, BrussSM13}
However, in our case simply increasing the interaction strength between the vertical binding sites to make protofilaments stiffer does not solve the twist issue 
because the self-assembly of free monomers with large interaction strengths results in kinetically trapped clusters instead of tubules.
In order to overcome this hurdle, we introduced a lock-and-key mechanism into our model monomers for the vertical interactions.
In this manner the control of the tubule self-assembly was substantially improved and our assembly simulations were able to achieve a good match between the tubule pitch and monomer chirality.

Our results reveal the importance of the vertical interaction strength being larger than the lateral strength in microtubules.\cite{VanBurenPNAS02}
This difference is necessary to prevent twist which would yield a wide range of structures that would not be biologically functional.
The simulations also provide new insight into the {\em initial} assembly dynamics of microtubules; most experimental work has studied the growth of microtubule ends but not the initial nucleation of microtubules.
Because the vertical interaction is stronger, the tubulin dimers first form protofilaments via vertical binding and the protofilaments subsequently bind laterally into curved sheets that close to form tubules.
This route generally enhances the structural control of assembled tubules and points to the importance of precise control
of the sheet-closing event.
Finally, we emphasize that controlling the assembly of tubules to form a specific number of protofilaments and helicity is highly nontrivial and requires many features in the monomer. 
It is thus not surprising that in cells additional constraints are imposed 
by other molecules (e.g., $\gamma$-tubulin) to achieve the degree of control observed for microtubules.

\section*{Acknowledgments}
This research was supported by the U.S. Department of Energy, 
Office of Basic Energy Sciences, Division of Materials Sciences 
and Engineering under Award KC0203010.
Sandia National Laboratories is a multi-program laboratory 
managed and operated by Sandia Corporation, 
a wholly owned subsidiary of Lockheed Martin Corporation, 
for the U.S. Department of Energy's National Nuclear Security Administration 
under contract DE-AC04-94AL85000.

%

\onecolumngrid

\setcounter{figure}{0}
\renewcommand{\thefigure}{S\arabic{figure}}

\section*{Supplementary Material of ``Self-Assembly of Chiral Tubules''}

\begin{figure}[htb]
\centering
\includegraphics[width=6.5in]{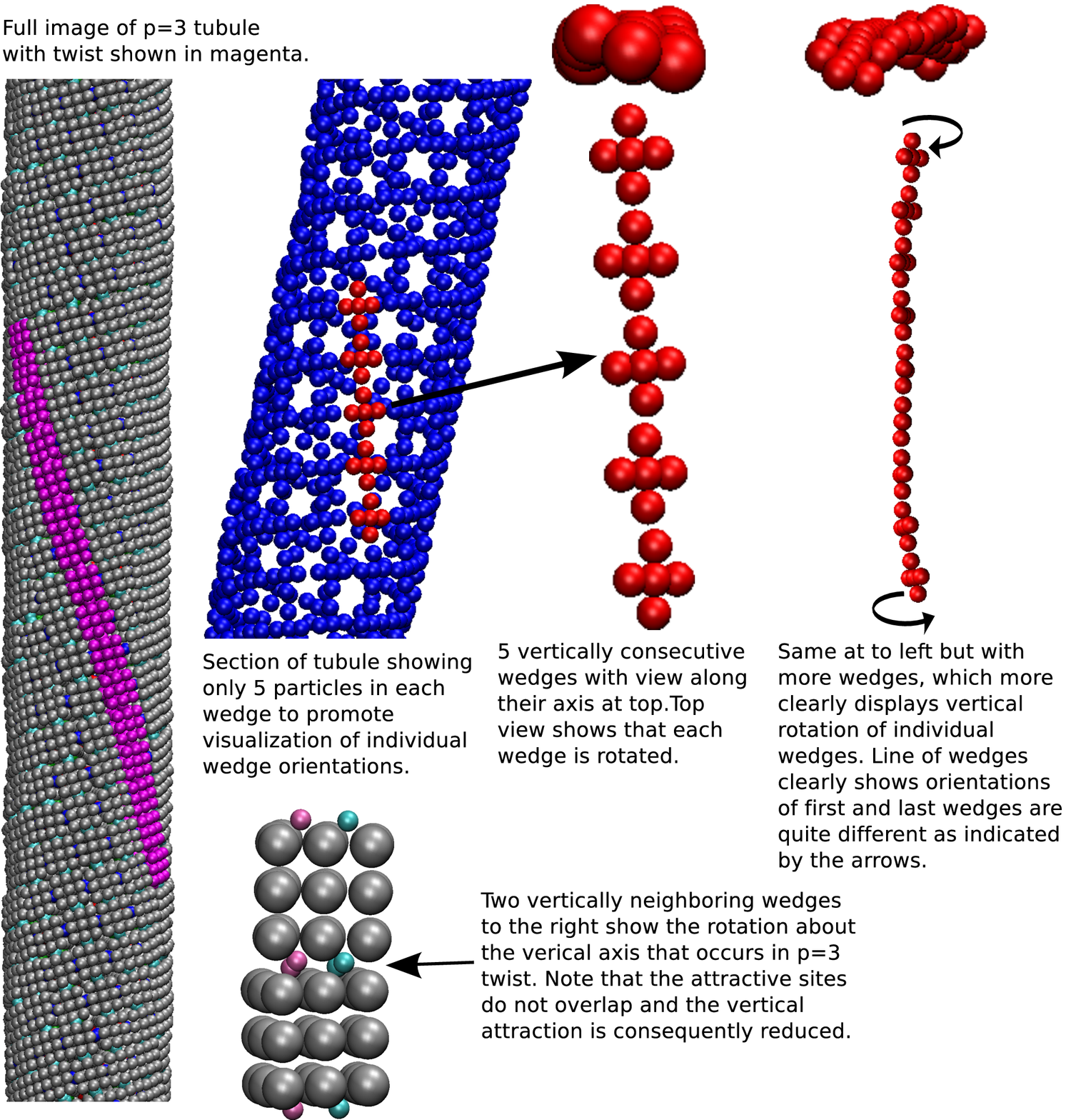}
\caption{
An equilibrated 13\_3 tubule of $M_0$ monomers with twisted protofilaments
shows the rotation of wedges along a protofilament.
}
\label{Fig_twist_p3}
\end{figure}

Using a 13\_3 tubule of $M_0$ monomers as an example, 
Fig.~\ref{Fig_twist_p3} illustrates the limits of twist deformation of protofilaments
when the helicity of a tubule does not match the
the chirality of monomers.
Because of the curvature of a tubule surface, 
a twisted protofilament has to curl around the central axis of the tubule,
which in turn requires the rotation of wedge monomers along the protofilament and
introduces an offset between the vertical binding sites 
between two stacking neighboring wedges.
If the required offset is too large, then the tubule will have a much higher energy than
the one with helicity that matches the chirality of monomers,
and the tubule will be energetically unfavored.
Therefore, the key to suppress the rotation of wedges 
and thus the twist of protofilaments
is to have a large $A_V$.

\begin{figure}[htb]
\centering
\includegraphics[width=3.25in]{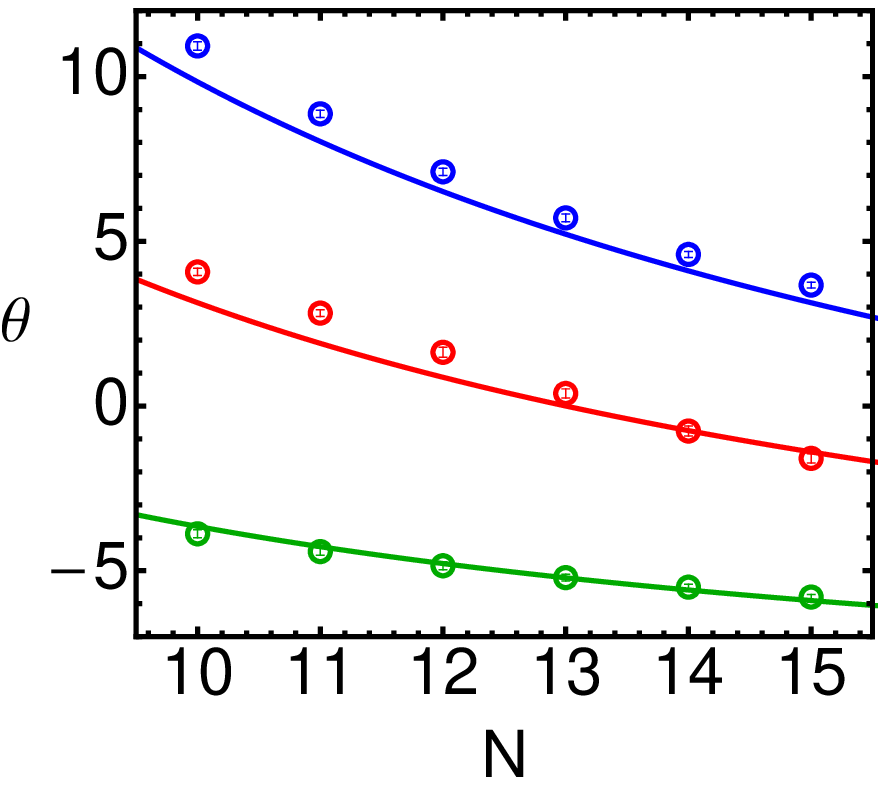}
\caption{
The twist angle $\theta$ of protofilaments in tubules
formed by $M_2^{LK}$ monomers and 
with various number ($N$) of protofilaments
and pitches: $p=2$ (red), $p=1$ (green), 
and $p=3$ (blue) at $A_L=3.0$ and $A_V=6.3$.
Symbols are simulation results 
with error bars comparable or smaller than the symbol size.
Lines represent the corresponding 
predictions of Eq.~(1) of the main text 
with $h=3\sigma$ and $w=2.53\sigma$.
}
\label{Fig_angle_VLKS2}
\end{figure}

Equation (1) of the main text describes 
the twist deformation of protofilaments for all tubules
built from any monomers that we have studied.
One more example is included in Fig.~\ref{Fig_angle_VLKS2}
for tubules formed by $M_2^{LK}$ monomers.
In general, it is expected that Eq.~(1) of the main text is applicable 
to all tubular structure made out of identical discrete
building blocks.

\begin{figure}[htb]
\centering
\includegraphics[width=6.5in]{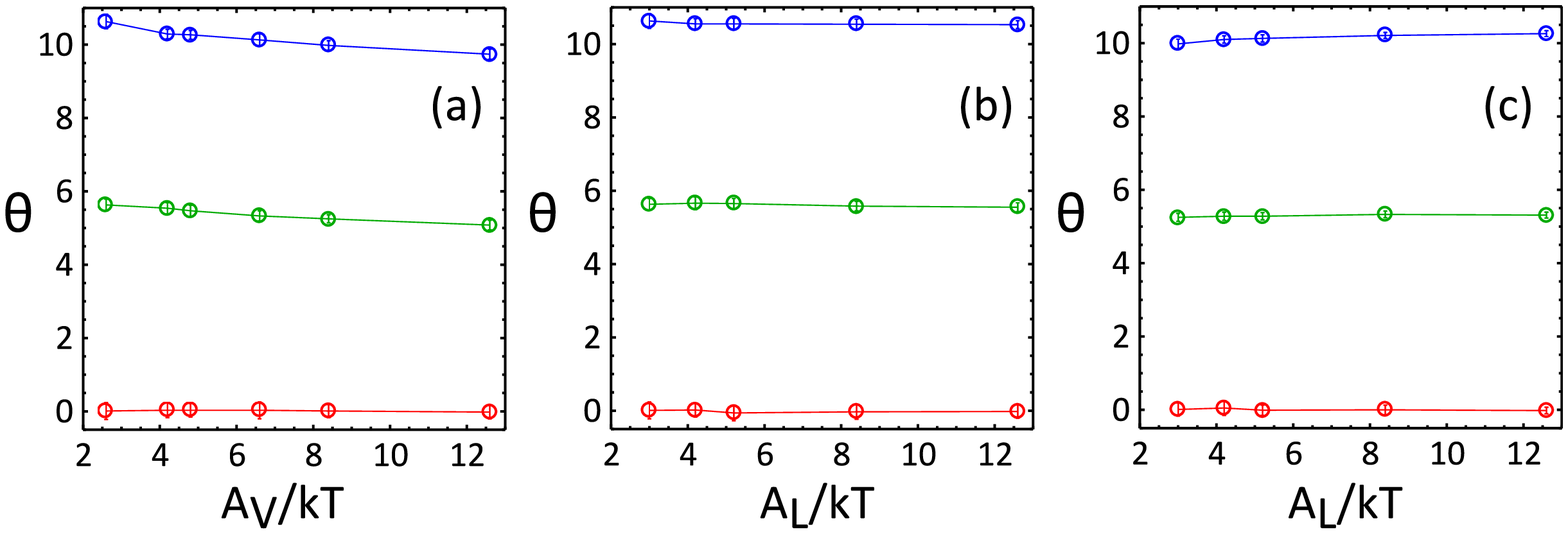}
\caption{
$\theta$ vs. interaction strength:
(a) $A_L=3.0$ and $A_V$ is varied;
(b) $A_V=2.6$ and $A_L$ is varied;
(c) $A_V=8.4$ and $A_L$ is varied.
Data are for $13\_p$ tubules built from $M_0$ monomers
with: $p=0$ (red/bottom), $p=1$ (green/middle), and $p=2$ (blue/top).
Lines are guides to the eye.
}
\label{Fig_angle_interaction}
\end{figure}

We study the effects on the tubule twist deformation 
of both lateral and vertical binding interactions.
One set of results is included in Fig.~\ref{Fig_angle_interaction}, which shows that
the twist angle $\theta$ is insensitive to either $A_L$ or $A_V$.
This is not surprising, because $\theta$ is mainly determined by the geometric features
of the building blocks, as expressed in Eq.~(1) of the main text.
However, the small changes of $\theta$ with $A_L$ or $A_V$ are still noteworthy.
On the one hand, $\theta$ slightly decreases as $A_V$ is increased while $A_L$ is fixed,
indicating that the twist deformation is slightly reduced at a larger $A_V$.
This trend can be understood on the basis that a large $A_V$ makes the protofilaments stiffer and helps reduce the offset
between the vertical binding sites of two stacking wedges in a protofilament 
(see Fig.~\ref{Fig_twist_p3}), and a smaller offset 
leads to a more gently twisted protofilament and thus a smaller twist angle.
On the other hand, $\theta$ essentially remains unchanged as $A_L$ is varied
at a small fixed $A_V$, or increases slowly with an increasing $A_L$ at a large fixed $A_V$,
which indicates that the twist deformation is slightly enhanced with a large $A_L$.
The underlying physics is that a larger $A_L$ implies a stronger adhesion between neighboring protofilaments,
which favors a twisted packing of protofilaments. 
The similar trends in $\theta$ vs.
$A_L$ and $A_V$ were observed
for all tubules that we have built with our monomers
(achiral/chiral, with/without lock-and-key vertical binding).

\begin{figure}[htb]
\centering
\includegraphics[width=5in]{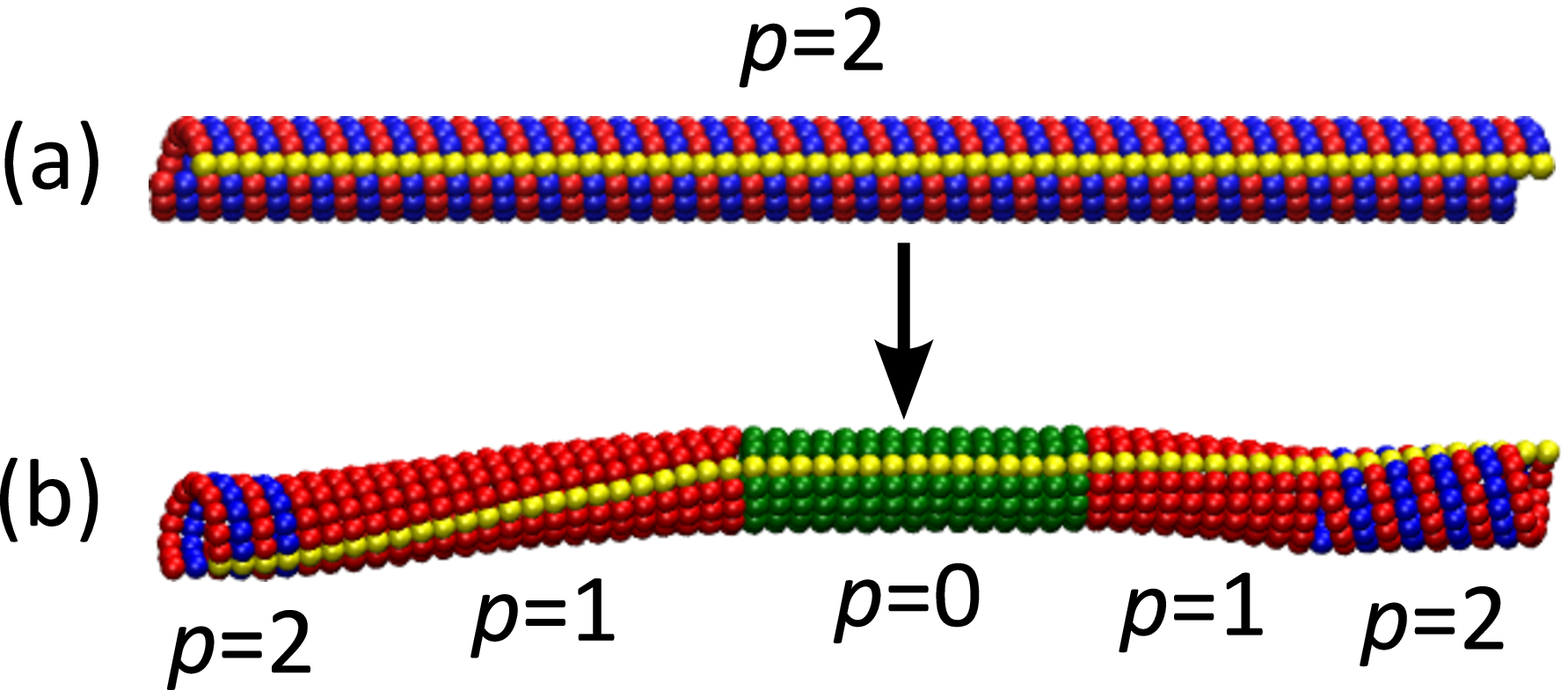}
\caption{
(a) A pre-built 13\_2 tubule of $M_0$ monomers that contains straight protofilaments.
(b) At $A_V=4.8>A_L=3.0$, the tubule in (a) evolves into a state with 
multiple pitches ($p=0$, 1, and 2) and
the protofilaments are twisted in portions with $p\ne 0$.
}
\label{Fig_pretubule_transition}
\end{figure}

Results on $\theta$ in Fig. 4 of the main text, Fig.~\ref{Fig_angle_VLKS2},
and Fig.~\ref{Fig_angle_interaction} were obtained with tubules starting with
protofilaments twisted according to Eq.~(1) of the main text.
However, if the starting state has straight protofilaments, 
then the pitch of tubules can change, especially when $A_V>A_L$. 
An example is shown in Fig.~\ref{Fig_pretubule_transition}
for a prebuilt 13\_2 tubule of $M_0$ monomers.
Here the starting tubule has straight protofilaments and 
$A_V=4.8> A_L=3.0$. The tubule quickly transforms into a hybrid structure 
of 13\_0, 13\_1, and 13\_2 tubules.
However, if we ran the simulation at $A_L > A_V$ with the same starting 
configuration where protofilaments are straight,
then the tubule stays at 13\_2, but ends up with twisted protofilaments 
(see Fig. 2 of the main text for the case $A_L=4.2>A_V=2.6$), 
of which the twist angle is consistent with Eq.~(1) of the main text.
The example in Fig.~\ref{Fig_twist_p3} is for a tubule with $p=3$ under 
$A_L=4.2 > A_V=2.6$. In that case the same final state with twisted protofilaments 
is achieved even by 
tubules starting with straight protofilaments, in contrast to the case $A_V>A_L$.
However, these tubules are still meta-stable, 
though when $A_L>A_V$ the depths of the local minima are increased
and their meta-stability is enhanced compared with the $A_V>A_L$ case.

\begin{figure}[htb]
\centering
\includegraphics[width=4in]{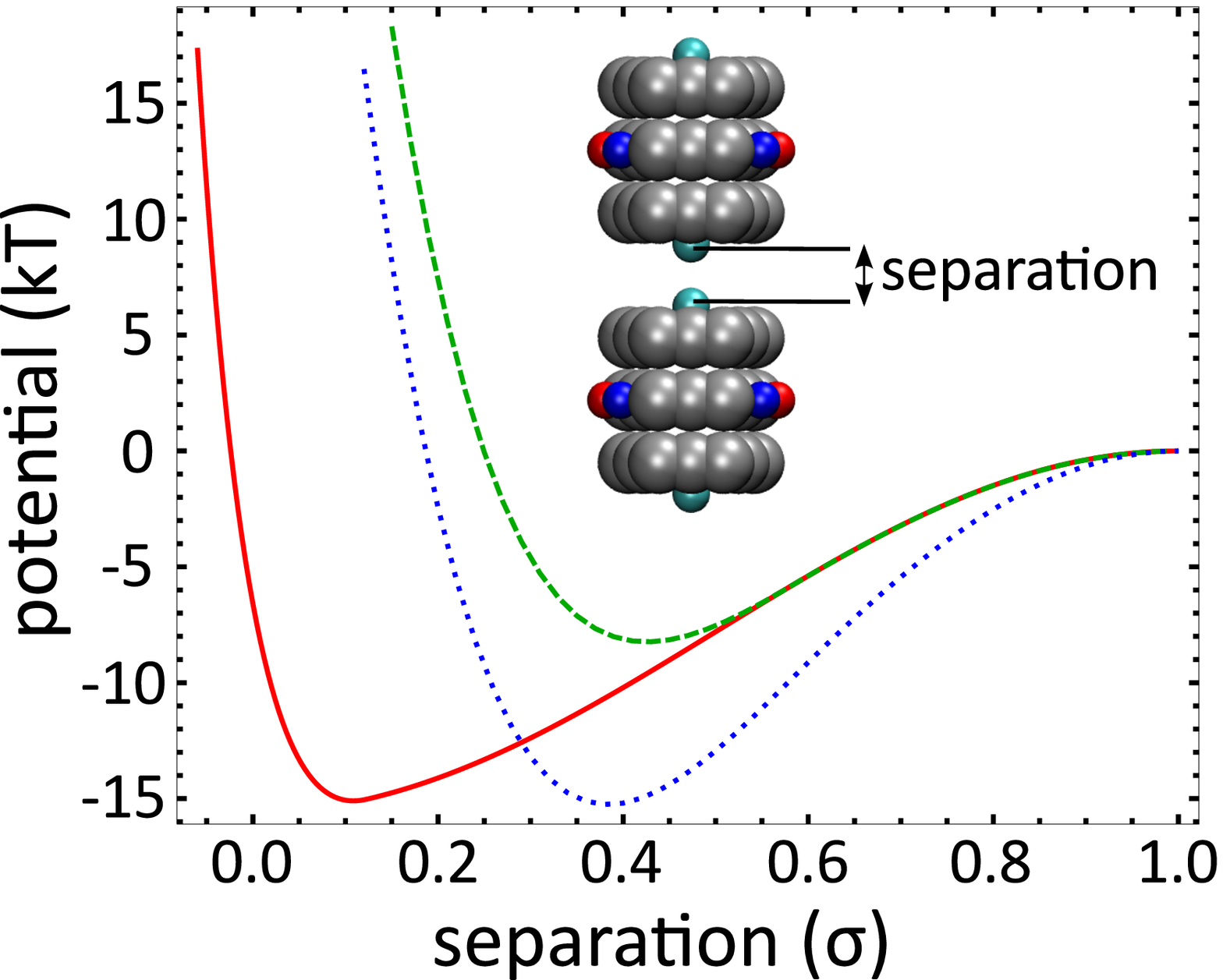}
\caption{
The potential energy as a function
of separation for two monomers binding vertically.
Three curves are for two $M_0$ monomers (inset picture) at $A_V=3.9$ (red solid line),
two $M_0^{LK}$ monomers at $A_V=3.9$ (green dashed line) and $A_V=6.6$ (blue dotted line).
}
\label{Fig_vertical_potential}
\end{figure}

Figure \ref{Fig_vertical_potential} helps understand why a larger $A_V$ is needed to initiate
the self-assembly of lock-and-key monomers. Here the total potential energy, which
is the sum of the repulsions between the gray core sites on two monomers
and the attractions between the colored attractive sites (see the Methods section
of the main text for more details on the wedge-wedge interactions),
is plotted as a function of the separation of wedges that bind at their vertical surfaces.
The two wedges are aligned vertically and the interaction energy is calculated as a function of separation.
The zero separation corresponds to 
the state in which the two pairs of vertical binding sites (the sites with the cyan and green color) of two monomers 
overlap.
Because of the repulsion between the gray core sites, the minimum of the potential energy occurs
at a positive separation, which is around 0.1$\sigma$ for the monomers without the lock-and-key configuration.
However, for the lock-and-key monomers (see Fig.~1 of the main text for the geometry), 
the location of the potential minimum shifts to a larger separation (about $0.4\sigma$ for the
lock-and-key monomers studied in this paper).
The underlying reason is as follows.
First, the gray core sites of two lock-and-key monomers start
to interact at a larger separation compared with the case without lock-and-key
and the interactions are purely repulsive.
Then at a given separation the total repulsion between two lock-and-key monomers
is always stronger than that between two monomers without lock-and-key,
while the total attraction between monomers is 
the same in the two cases at the same separation.
As a consequence, at a given $A_V$ the depth of the potential well of two vertically bound monomers
is more shallow for the lock-and-key monomers and the location of its minimum 
moves to a more positive separation,
which is clearly seen from the comparison at $A_V=3.9$ in Fig.~\ref{Fig_vertical_potential}.
To compensate for this reduction, a larger $A_V$ is needed for the lock-and-key monomers to 
achieve the same total attraction when two wedges bind vertically.
For example, the well depth of the vertical binding of two $M_0^{LK}$ monomers at $A_V=6.6$ 
is close to that of two $M_0$ monomers at $A_V=3.9$, as shown in Fig.~\ref{Fig_vertical_potential}.
We also find that when the well depth is similar, the curvature of the potential energy around its minimum becomes larger 
after the introduction of the lock-and-key vertical binding mechanism, which indicates that
the potential becomes stiffer and the fluctuations in the vertical bonds are reduced.

Tables~\ref{table_shift} and \ref{table_notch} summarize all results from assembly simulations
starting with free monomers.
As noted in the main text, $N\_p$ stands for tubules containing 
$N$ protofilaments organized with pitch $p$.
Tubules with $p=c$ are shown in bold, where $c$ is the chirality of monomers.
The mismatch between $p$ and $c$ frequently occurs for monomers without lock-and-key
vertical binding mechanism, as shown in Table~\ref{table_shift}.
However, the mismatch can be reduced or even suppressed using monomers with vertical lock-and-key
binding, especially when $A_V>A_L$ as shown in Table~\ref{table_notch}.
In general, for lock-and-key monomers with $0\le c \le 2$, $N\_p$ tubules with $p=c$
dominate when $A_V>A_L$, while $N\_p$ tubules with $p=c \pm 1$ dominate when $A_L>A_V$.
For $M_3^{LK}$ monomers, $N\_3$ tubules are only found when $A_V>A_L$.
For $M_3$ monomers, tubules with $p=2$ are the most frequently assembled structures, 
no matter $A_V>A_L$ or $A_L>A_V$.

\begin{table}[h]
\centering
\begin{tabular}{|c|c|>{\centering}p{3.2cm}<{\centering}|>{\centering}p{3.2cm}|
p{3.2cm}<{\centering}|p{3.2cm}<{\centering}|} \hline
 $A_L$ & $A_V$ & $M_0$ & $M_1$ & $M_2$ & $M_3$\\ \hline
3.0 & 3.9 & {\bf 11\_0, 12\_0,} 11\_1, 12\_1 & {\bf 11\_1, 12\_1,} 11\_2 &
 {\bf 11\_2, 12\_2} & 11\_2\\ \hline
4.2 & 2.6 & 12\_1, 11\_2, 12\_2 & 12\_0, {\bf 12\_1,} 11\_2 &
 12\_0, 13\_0, 11\_1, 12\_1, {\bf 12\_2, 13\_2} & 12\_1, 11\_2, 12\_2, {\bf 12\_3} \\ \hline
\end{tabular}
\caption{Tubule formation by $M_c$ monomers}
\label{table_shift}
\end{table}  

\begin{table}
\centering
\begin{tabular}{|c|c|>{\centering}p{3.2cm}<{\centering}|>{\centering}p{3.2cm}|
p{3.2cm}<{\centering}|p{3.2cm}<{\centering}|} \hline
$A_L$ & $A_V$ & $M_0^{LK}$ & $M_1^{LK}$ & $M_2^{LK}$ & $M_3^{LK}$\\ \hline
3.0 & 6.3 & {\bf 11\_0, 12\_0} & {\bf 11\_1, 12\_1} 
 & {\bf 11\_2, 12\_2} & 11\_2, 12\_2, {\bf 11\_3, 15\_3}\\ \hline
3.0 & 6.0 & No assembly & {\bf 11\_1, 12\_1} 
 & {\bf 11\_2, 12\_2}  & 11\_2, 12\_2\\ \hline
3.3 & 6.3 & Clusters & 12\_0, {\bf 11\_1}  & {\bf 11\_2} 
 & 11\_2,  {\bf 13\_3}\\ \hline
3.3 & 6.0 & {\bf 11\_0, 12\_0,} 11\_1, 12\_1 & 14\_0, {\bf 11\_1, 12\_1} 
 & 12\_1, {\bf 11\_2, 12\_2} & 11\_2, 12\_2, {\bf 12\_3}\\ \hline
3.3 & 5.7 & No assembly & {\bf 12\_1} & 11\_1, {\bf 11\_2, 12\_2, 13\_2} 
 & 11\_2, 12\_2, 13\_2, {\bf 11\_3}\\ \hline
3.6 & 6.0 & {\bf 12\_0,} 12\_1 & {\bf 12\_1} & 13\_1, {\bf 11\_2, 12\_2} &
 13\_2, {\bf 12\_3, 14\_3}, 13\_4\\ \hline
3.6 & 5.7 & {\bf 11\_0, 12\_0} & {\bf 11\_1, 12\_1} 
 & 11\_1, 12\_1, {\bf 12\_2} & 12\_1, 11\_2, 12\_2, 13\_2\\ \hline
3.6 & 5.4 & {\bf 12\_0,} 12\_1 & {\bf 11\_1, 12\_1} 
 & 11\_1, {\bf 11\_2, 12\_2} & 12\_2, {\bf 11\_3}\\ \hline
3.9 & 5.1 & {\bf 12\_0}, 12\_1 & 12\_0, {\bf 11\_1, 12\_1, 13\_1}, 11\_2 
 & 12\_1, {\bf 12\_2, 13\_2} & 12\_1, 12\_2, {\bf 12\_3}\\ \hline
4.4 & 4.2 & {\bf 12\_0}, 12\_1, 13\_1 & 12\_0, 14\_0, {\bf 11\_1, 12\_1, 13\_1} 
 & 12\_0, 13\_0, 12\_1, 13\_1, {\bf 12\_2,} 11\_3 & 11\_1, 12\_1, 11\_2, 13\_2\\ \hline
4.4 & 3.9 & {\bf 13\_0}, 13\_1 & 12\_0, {\bf 12\_1, 13\_1} 
 & 12\_0, 11\_1, 13\_1, {\bf 11\_2, 12\_2} & 11\_0, 12\_0, 12\_1\\ \hline
4.8 & 3.6 & {\bf 13\_0}, 12\_1 & 12\_0, 13\_0, 14\_0, {\bf 12\_1}, 12\_2 
 & 12\_0, 13\_0, 14\_0, 12\_1, 13\_1, {\bf 12\_2} &11\_0, 12\_0, 11\_1, 12\_1, 13\_1\\ \hline
\end{tabular}
\caption{Tubule formation by $M_c^{LK}$ monomers.}
\label{table_notch}
\end{table} 

\end{document}